\documentclass{edp-conf}
\usepackage{graphicx}
\def\uu{4U\,1608-52}
\def\PI{Paper I}
\def\apjl{ApJL}

\def\araa{ARA\&A}

\begin{document}

\TitreGlobal{SF2A 2004}

\title{Rapid variability of the kHz-QPO frequency in \uu}
\author{Paltani, S.}\address{Laboratoire d'Astrophysique de Marseille, BP 8, 13376 Marseille cedex 12, France}
\author{Barret, D.}\address{CESR, 9 avenue du Colonel Roche, 31028 Toulouse Cedex 4, France}
\author{Olive, J.-F.$^2$}
\author{Skinner, G. K.$^2$}
\runningtitle{Rapid variability of the kHz-QPO frequency in \uu}
\setcounter{page}{237}
\index{Paltani, S.}
\index{Barret, D.}
\index{Olive, J.-F.}
\index{Skinner, G. K.}

\maketitle
\begin{abstract} We investigate the variability of the QPO frequency
  in \uu\ on very short time scales of a few seconds. We detect
  changes in frequency as fast as 0.5\,Hz/s, which is more than ten
  times the fastest change reported for this object. Using a structure
  function analysis, we are able to detect variability on time scales
  as short as $\sim 4$\,s. We do not detect any time asymmetry in the
  QPO-frequency history. These results have important consequences on
  the measurement of the QPO coherence.
\end{abstract}
%
\section{Introduction}
Fourier analysis has shown that the X-ray emission of low-mass X-ray
binaries (LMXBs) often presents characteristic
quasi-periodic-oscillation (QPO) peaks at frequencies up to $\sim 1$
kHz (van der Klis 1989,
2000)\nocite{vdKl-1989-QuaPer,vdKL-2000-MilOsc}.  The kHz QPOs are of
particular interest, as they are a probable signature of physical
mechanisms taking place in the inner parts of the accretion disk,
where strong-field effects of gravity become important.  The physical
origin of the QPOs is still a matter of debate, and a detailed
understanding of the phenomenon is therefore necessary. Berger et al.
(1996)\nocite{BergEtal-1996-DisQua} analyzed the variation of the QPO
frequency in a Rossi-XTE observation and observed a 40\,Hz change in
frequency in about 1000\,s.  They also found evidence for a high QPO
coherence of the order of 200.  Building upon these results, Barret et
al. (2005)\nocite{BarrEtal-2004-HigQua} (hereafter \PI) have recently
made important progress in the determination of the width of the kHz
QPO in \uu. In \PI\ they found evidence that the variability of the
QPO frequency could be quite significant, even on short time scales.
In this work, we analyze the temporal properties of the QPO frequency
in \uu, focussing on the shortest possible time scales allowed by the
signal.

\section{Data}

We use observations of \uu\ from the Rossi-XTE public archive. These
data have already been analyzed in Berger et al.
(1996)\nocite{BergEtal-1996-DisQua} and M\'endez et al. (1998a, 1998b,
1999)\nocite{MendEtal-1998-DisSec,MendEtal-1998-KilQua,MendEtal-1999-DepFre},
as well as in \PI. We selected $\sim 1000$ seconds from an observation
segment of \uu\ performed on March 3rd, 1996 20:54. This data set has
been chosen because the QPO is particularly strong and highly
coherent, and its frequency has been found in \PI\ to remain in a
narrow frequency range.

\section{Analysis methods}
\subsection{Rayleigh's test as an estimator of the PSD}
We consider the individual photons, without building a light curve.
The observed signal is therefore a sum of Dirac $\delta$ functions:
\begin{equation}
S(t) = \sum_i \delta ( t - t_i )
\end{equation}
where the $t_i$'s are the time tags of the individual photons. The
Fourier transform and the power-spectrum density (PSD) of this signal
can be easily obtained:
\begin{equation}
\mathrm{FT} ( S ) = \sum_i \exp ( 2 \pi \nu \iota t_i )
\hspace*{5mm} \Rightarrow \hspace*{5mm}
\mathrm{PSD} ( S ) = \| \sum_i \exp ( 2 \pi \nu \iota t_i ) \| ^2
\end{equation}
This is exactly the expression of Rayleigh's test (e.g., Fisher
1993)\nocite{Fish-1993-StaAna}. We shall use this expression to
estimate the PSD of the light curve of \uu, in order to avoid any loss
of information due to the binning of the photons.

\subsection{QPO-frequency determination}
We determine the variation of the QPO frequency using a simple method:
We calculate the PSD over time intervals as short as a few seconds,
and smooth it using a Gaussian; we then search the maximum of the
smoothed PSD.  This method has been validated using simulations of a
QPO at a known frequency with similar amplitude and width (see \PI).
We find that the accuracy of the QPO frequency determination depends
little on the width of the Gaussian, provided that it is significantly
wider than the natural resolution of the Fourier transform, i.e.\ 
$\Delta\nu\sim 1/T$, where $T$ is the duration of the time interval.
We used a Gaussian with a FWHM of 3\,Hz for time intervals from 1 to
10\,s.

Statistical fluctuations may sometimes produce spurious peaks which
exceed the strength of the QPO. To eliminate these peaks, we first
calculate the PSDs from $\nu=100$ up to $\nu=2000$\,Hz over 10s time
bins. With such time intervals the measured QPO frequency always fall
within a few Hz from the curve determined in \PI. When we shorten the
time bin, spurious peaks appear. To get rid of these peaks, we first
restrict the range of investigated frequencies to $\nu=810-850$\,Hz,
and reject QPO-frequency measurements that fall farther than 5\,Hz
away from the 10s-bin frequency. Using this procedure we eliminate
about 2\% of the peaks when using 4s bins. This figure increases to
11\% for 2s bins, and to 25\% for 1s bins.

\section{Results}
\label{sec:results}
\subsection{QPO frequency as a function of time}
\begin{figure}[tb]
  \centering
  \includegraphics[width=5cm]{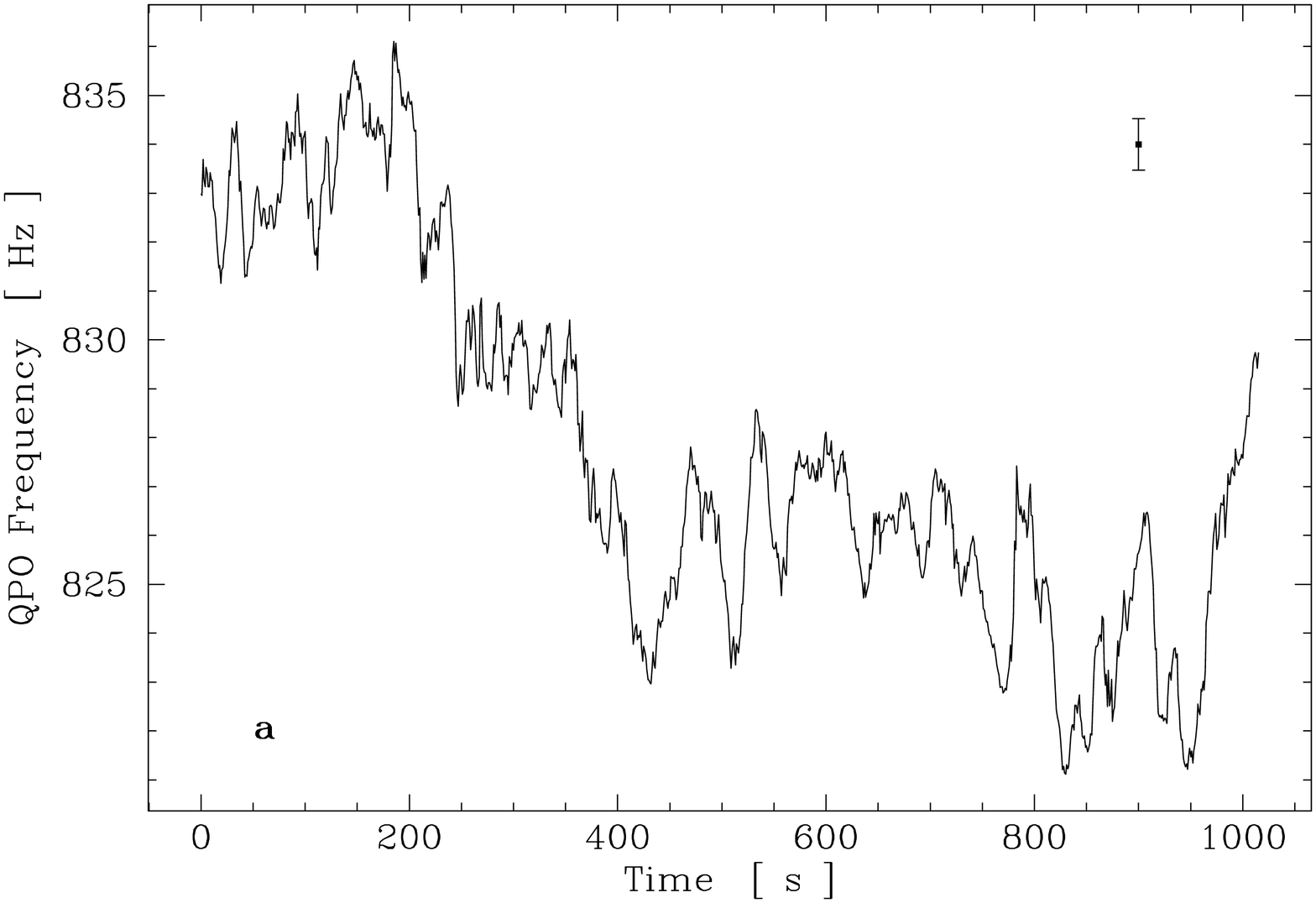}\hspace*{8mm}
  \includegraphics[width=5cm]{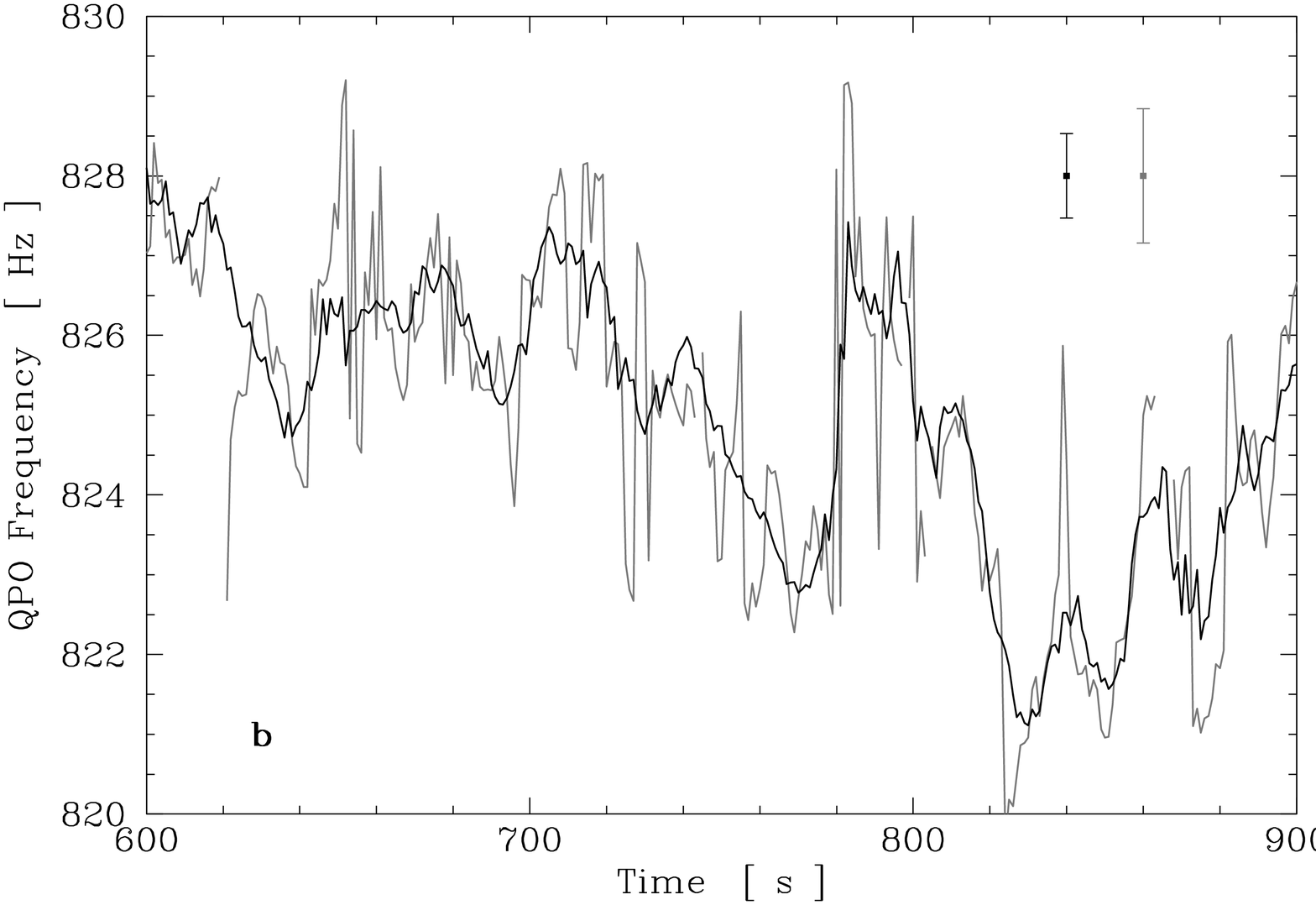}
  \caption{{\bf (a)} Evolution of the QPO frequency calculated
    every 1\,s over 10s bins. {\bf (b)} Zoom on the QPO-frequency
    evolution in 10s (black curve) and 4s (grey curve) bins. In each
    panel the average uncertainties are shown in the upper right
    corner.}
  \label{figure:qpo}
\end{figure}
Fig.~\ref{figure:qpo}a shows the evolution of the QPO frequency as a
function of time, $\nu_{\mathrm{QPO}}(t)$, for the full segment of
data analyzed here. The QPO frequency has been determined every 1
second using 10\,s time bins. Very rapid changes of frequency as large
as 0.5\,Hz/s are observed, both towards higher and lower frequencies.
A Kolmogorov-Smirnov test on the distributions of the positive and
negative frequency changes does not show any evidence of time
asymmetry.  Fig.~\ref{figure:qpo}b zooms in on a portion of the data,
on which we superimpose $\nu_{\mathrm{QPO}}(t)$ calculated in 10s and
4s bins.  While the curves do follow each other quite well,
$\nu_{\mathrm{QPO}}(t)$ with 4\,s bins show much more prominent
structures that it seems difficult to ascribe to the increased
uncertainties only.

\subsection{Time-series properties}
\begin{figure}[tb]
  \centering
  \includegraphics[width=5cm]{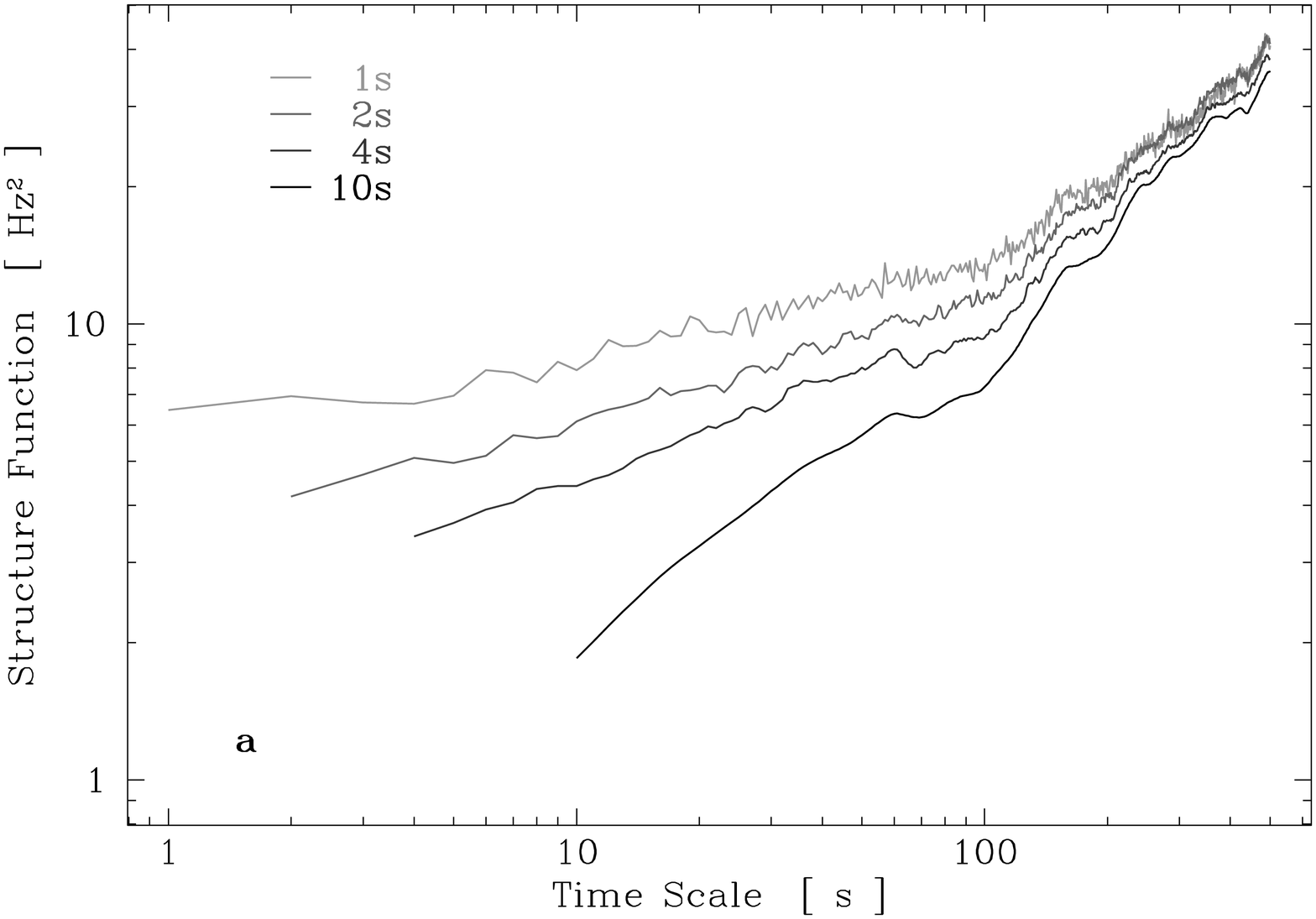}\hspace*{8mm}
  \includegraphics[width=5cm]{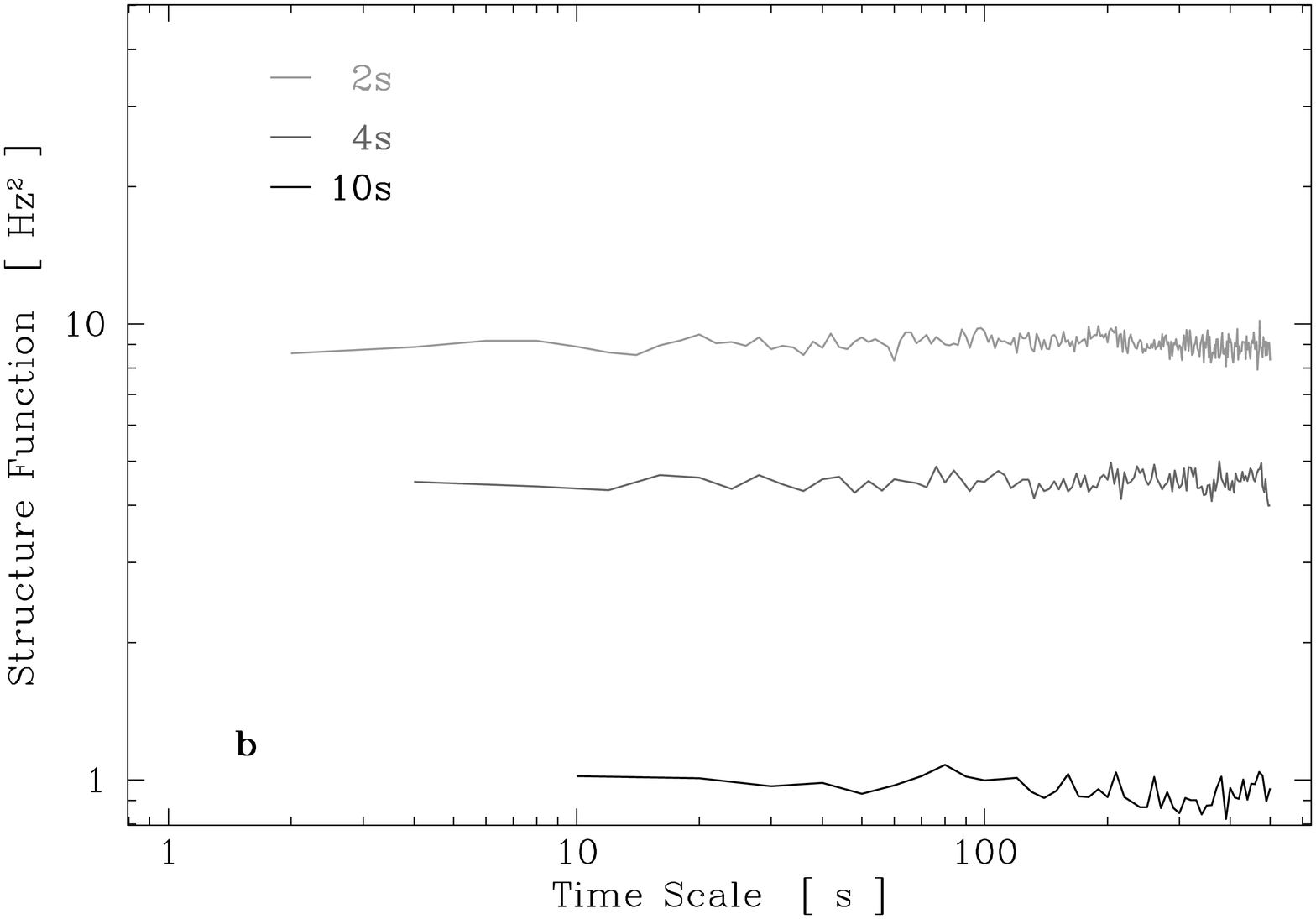}
  \caption{SFs of $\nu_{\mathrm{QPO}}(t)$ for the observed {\bf
      (a)} and simulated {\bf (b)} data for different bin sizes. The
    bin size increases from top to bottom.}
  \label{figure:sf}
\end{figure}
To investigate and quantify the variability of the QPO frequency, we
compute the structure functions (SF) of $\nu_{\mathrm{QPO}}(t)$ for
the data and for simulations. The SF of a time series (e.g., Paltani
1999\nocite{Palt-1999-ConBll}) measures the amount of variability as a
function of the time scale $\tau$.  It is flat for a
white-noise-dominated time series.  The SF increases with $\tau$ as
long as the time series does contain some variability at the
corresponding time scale $\tau$.  Fig.~\ref{figure:sf}a shows the SFs
of $\nu_{\mathrm{QPO}}(t)$ for different time bins. For comparison
Fig.~\ref{figure:sf}b shows the same SFs for the simulations.  These
SFs are dominated by the uncertainty in the frequency determination,
which shows that our method does not produce any artefact. For the
\uu\ data, the horizontal part is present in the SF of
$\nu_{\mathrm{QPO}}(t)$ calculated in 1\,s bins only. Changes in the
QPO frequency on time scales as short as $\sim 4$\,s are therefore
detected. The differences in the SFs are the result of the smoothing
and of the decrease in the uncertainty introduced by the use of larger
bins.

\section{Conclusion}
We observe the presence of an erratic variability of the QPO frequency
in \uu, with evidence of changes in QPO frequency as large as
0.5\,Hz/s. This is already more than ten times faster than previously
reported by Berger et al. (1996)\nocite{BergEtal-1996-DisQua}. Faster
frequency changes seem present in $\nu_{\mathrm{QPO}}(t)$ calculated
with shorter time bins. The signal is strong enough, and the method
developed here sufficiently accurate, that a minimum variability time
scale of $\sim 4$\,s is detected. We did not find any evidence for
asymmetry in $\nu_{\mathrm{QPO}}(t)$.  These rapid and erratic
frequency changes have consequences for the measurement of QPO
coherence. A blurring of the QPO profile will indeed take place if too
long time bins are used, leading to an underestimation of the true QPO
coherence.  As discussed in \PI, the high coherence of the signal
places very strong constraints on QPO models.  It is therefore
important to measure it on very short time scales.




\end{document}